\documentclass[pra,amsmath,amssymb,superscriptaddress,showpacs,twocolumn]{revtex4-1}
\usepackage{CJK}
\usepackage{graphicx}
\usepackage{amsmath}
\usepackage{bm}
\usepackage{graphicx}
\usepackage{dcolumn}
\usepackage{bm}% bold math
\usepackage{colordvi}
\usepackage{color}
\usepackage{ulem}
\newcommand{\beq}{\begin{equation}}
\newcommand{\eeq}{\end{equation}}
\newcommand{\beqa}{\begin{eqnarray}}
\newcommand{\eeqa}{\end{eqnarray}}

\def\ra{\rangle}
\def\la{\langle}

\begin{document}

\title{Quantum state engineering of spin-orbit coupled ultracold atoms in a Morse potential}

\author{Yue Ban}
%\email{yban@shu.edu.cn}
\affiliation{Department of Electronic Information Materials, Shanghai University, 200444 Shanghai, People's Republic of China }

\author{Xi Chen}
%\email{xchen@shu.edu.cn}
\affiliation{Department of Physics, Shanghai University, 200444 Shanghai, People's Republic of China }

\author{J. G. Muga}
\affiliation{Departamento de Qu\'{\i}mica-F\'{\i}sica, UPV/EHU, Apdo 644, 48080 Bilbao, Spain}
\affiliation{Department of Physics, Shanghai University, 200444 Shanghai, People's Republic of China }

\author{E. Ya Sherman}
\affiliation{Departamento de Qu\'{\i}mica-F\'{\i}sica, UPV/EHU, Apdo 644, 48080 Bilbao, Spain}
\affiliation{IKERBASQUE, Basque Foundation for Science, 48011 Bilbao, Spain}

%\author{J. G. Muga}
%\affiliation{Departamento de Qu\'{\i}mica-F\'{\i}sica, UPV/EHU, Apdo 644, 48080 Bilbao, Spain}
%\affiliation{Department of Physics, Shanghai University, 200444 Shanghai, People's Republic of China }

\date{\today}
\begin{abstract}
Achieving full control of a Bose-Einstein condensate can have valuable applications in metrology, quantum information processing, and quantum condensed matter physics. We propose protocols to simultaneously control the internal (related to its pseudospin-1/2) and motional (position-related) states of a spin-orbit-coupled Bose-Einstein condensate confined in a Morse potential. In the presence of synthetic spin-orbit coupling, the state transition of a noninteracting condensate can be implemented by Raman coupling and detuning terms designed
by invariant-based inverse engineering. The state transfer may also be driven by tuning the direction of the
spin-orbit-coupling field and modulating the magnitude of the effective synthetic magnetic field. The results
can be generalized for interacting condensates by changing the time-dependent detuning to compensate for the
interaction. We find that a two-level algorithm for the inverse engineering remains numerically accurate even if
the entire set of possible states is considered. The proposed approach is robust against the laser-field noise and
systematic device-dependent errors.%
\end{abstract}
\maketitle
\section{Introduction}
%
%
%
%IMPORTANCE OF CONTROL QUANTUM SYSTEMS
Coherent high-fidelity control of quantum systems is a fundamental task in many areas of atomic, molecular, optical,
and condensed matter physics. Algorithms of such a control can be applied in metrology, interferometry, and quantum
information processing. Specifically, achieving fast and stable manipulation of ultracold ensembles of bosonic and fermionic
atoms by driving the system from an initial to a target state with high fidelity has been a major research goal during the
past two decades.

%EXAMPLES:
%Ultracold atoms in a single trap, double wells and optical lattices can be manipulated by adjusting
%laser beams and magnetic fields to control the potentials and interaction precisely.
Motional state control of localized atoms in particular, can be achieved by techniques inspired by
trapped-ion technology, or via trap deformations
\cite{Schmiedmayer-JPB,multiplexing}.
Synthetic spin-orbit (SO) coupling has also been proposed
to control or measure the orbital motion of atoms \cite{SOC-tunneling,SOC-measurement}.

In recent years, laser control techniques have successfully produced synthetic SO coupling in ultracold ensembles of neutral atoms,  such as Bose-Einstein condensates (BEC) \cite{SOC-BEC1,SOC-BEC2} and fermions \cite{SOC-fermion1,SOC-fermion2,SOC-fermion3}. SO-coupled BECs (see the recent reviews \cite{review1,review2,review3}) allow for control of several tunable parameters.  The combination of tunable SO coupling with interatomic interactions leads to novel phenomena unprecedented in conventional
condensed matter physics. SO-coupled condensates have been used, for example, to study and control spin dynamics in
processes such as spin relaxation  spin relaxation \cite{spin-relaxation-K}, \textit{Zitterbewegung} \cite{Zitterbewegung-spinhalf,Zitterbewegung-spin1}, spin resonance, and the spin-Hall effect. Tunable Landau-Zener transitions in a spin-orbit coupled BEC were experimentally studied \cite{Landau-Zener-SOC}.

In this paper, we we study the control of the dynamics of an
SO-coupled BEC confined in a Morse potential by inverse engineering \cite{PRL104,STA}
the control parameters. In this analytically solvable potential, the level spacing decreases as the energy
approaches the continuous spectrum and its spatial asymmetry implies a displacement of the center of mass for transitions
between vibrational states.  In the configuration producing experimentally synthetic spin-orbit coupling \cite{SOC-BEC2}, the control of the internal states can be implemented by tuning the coupling of the atomic pseudospin to the laser field, similarly to the invariant-based inverse engineering for quantum control in quantum dot \cite{BanPRL}.
For example, the amplitude of the external synthetic magnetic field and the direction of the SO-coupling
field can be chosen as the tunable parameters to control simultaneously the internal state and the position transfer,
resulting from the effect of the synthetic SO coupling on the orbital motion.

The paper is organized as follows:  In Section II, we introduce the model, reduce it to the effective two-level system, and formulate the initial and the target states of the transfer. In Section III, time-dependent Raman coupling and Raman detuning are designed to control internal and motional states by invariant-based inverse engineering for a noninteracting
condensate. Here we demonstrate that the designed two-level algorithm is applicable and gives a high fidelity for a more complicated multilevel system as well.
In Section IV, the direction of the SO-coupling field and the level detuning are designed to achieve the state transfer. The inverse engineering method is
generalized here for an interacting BEC by using a simple ansatz of the state evolution on the Bloch sphere. The robustness of this protocol with respect to noise and systematic
errors is discussed in Sec. V, and a short summary is provided in Sec. VI.

\section{Model and Hamiltonian}
\begin{figure}[t]
\begin{center}
\scalebox{0.5}[0.5]{\includegraphics{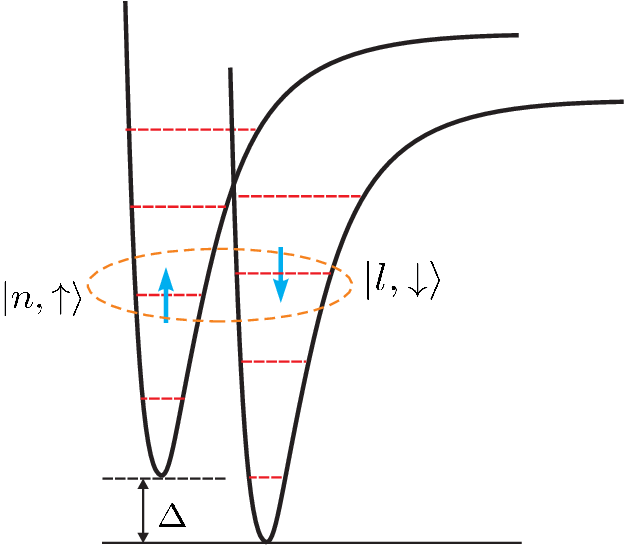}}
\caption{(Color Online) Schematic configuration of bosonic atoms trapped in a Morse-potential in the presence of SO coupling and the external effective magnetic field, by which degeneracy was eliminated and the energy gap between $|n,\uparrow \ra$ to $|l,\downarrow \ra$ is much less than the neighbouring orbital states. }
\label{schematic}
\end{center}
\end{figure}

We consider ultracold bosonic atoms trapped in a one-dimensional (1D) Morse potential \cite{Morse} of the form
\beqa
\label{U}
U(x) =  A \left( e^{-2 a x } - 2 e^{-a x } \right),
\eeqa
as shown in Fig. \ref{schematic} (a), where the characteristic parameters $A$ and $a$
%, in the unit of MHz and $\mu$m$^{-1}$,
have units of energy and inverse length, respectively. We assume that this potential is
independent of the relevant internal states. It can be tuned from an harmonic trap to a constant potential.
For large value of $a$, the potential is a good confining one, whereas it becomes constant $-A$ which is an effectively free particle, when $a \rightarrow 0$.
The Morse potential has been traditionally considered as a model for diatomic molecules, with coexisting unbounded and bound states. In atomic physics, the Morse potential can be produced with two evanescent light waves \cite{morse-potential}.
Due to its asymmetry, internal state control can be performed together with position displacements, which is a useful combination for interferometric applications so that each interferometer arm is subjected to  different effects.

The harmonic frequency is $\omega = \sqrt{2 A a^2 / M}$ near the minimum of the trap, for which the characteristic length $l_{\textrm{c}} = \sqrt{\hbar /M \omega}$ may be defined.
Hereafter we shall use dimensionless variables based on the unit of length $a^{-1}$,
the energy unit $a^2\hbar^{2}/M$, the time unit $ M / (a^2\hbar)$, and the velocity unit $\hbar a/ M$, where $M$ is the mass of an atom, which we also take as mass.  This results in an effective Schr\"odinger equation with $M=\hbar=1$.
The normalized orbital states in $x-$coordinate representation have the form
%%%%%%
\beqa
\nonumber
\la x | n \ra = n! z^{\xi} \sqrt{\frac{2  \xi}{\gamma(n+1)\gamma(2\eta-n)}} \exp\left(-\frac{z}{2}\right) L_n^{2 \xi} (z),
\\
\label{wavefunction}
\eeqa
with eigenvalue $E_{n} = -  (\eta - n - 1/2)^2 /2$,
where $\gamma(n)$ is the Euler gamma function, $\eta = \sqrt{2 A} $, $\xi = \eta -n -1/2$, $z \equiv z(x) = 2 \eta \exp(- x)$, and $L$ is the Laguerre polynomial.
%%%
%%%

A cold-atom moving in a 1D Morse potential (in $x$ direction) with a Raman laser configuration pointing in $z$ direction
coupling two pseudo-spin internal states is described by the Hamiltonian \cite{SOC-BEC2}
\beqa
\label{Href}
\mathcal{H}_0 = \frac{p^2}{2 } + U(x) + \frac{\Omega(t)}{2} \sigma_x + \frac{\Delta(t)}{2} {\sigma}_z +\alpha p {\sigma}_z,
\eeqa
where $p$ is the momentum of the atom, $\sigma_z$ is the $z$ component of the Pauli matrix, $\alpha$ is
the SO coupling strength due to the Doppler effect,
$\Delta$ is the detuning from resonance (effective Zeeman splitting), and $\Omega$ the effective Rabi frequency of the two-level system (Raman coupling strength) \cite{SOC-BEC2}.
For $\Omega=0$ the Hamiltonian $\mathcal{H}_0$ has ``spin-up'' eigenstates  $|n, \uparrow \ra = e^{-i  \alpha x} | n \ra |\uparrow \ra$ (with eigenvalue $E_{n, \uparrow}$), and ``spin-down''
eigenstates $|l, \downarrow \ra = e^{i  \alpha x} | l \ra |\downarrow \ra$ (with eigenvalue $E_{l, \downarrow}$, $l = n+1$), as shown in Fig. \ref{schematic}, where $|\uparrow \ra$ and $|\downarrow \ra$ are the eigenspinors of $\sigma_z$. In the following, different schemes of inverse engineering are proposed to transfer the state from $|n,\uparrow \ra$ to $|l,\downarrow \ra$. It is assumed that $\Delta$ is of the order of $| E_n - E_l|$ so that the energy gap $\Delta E = | E_{n, \uparrow} - E_{l, \downarrow}|$ is much smaller than the gap between the neighboring orbital states. In this way, the designed external field and SO coupling will have negligible effect on other vibrational states.
In addition, operation times will satisfy $t_f \gg 1/\Delta E$ and excitations to other states will be neglected. With the above two assumptions, an effective two-level system can be established to describe the state transfer in the 1D Morse potential.

\section{Scheme 1: State transition by tunable Raman Coupling and detuning}
%Hamiltonian with synthetic SO coupling induced by two laser beams \cite{SOC-BEC2} is in the form,
%%%%%%%%%%%%%
%\beqa
%\label{H-nature}
%\mathcal{H} = \frac{p^2}{2 } + U(x) + \frac{\Omega(t)}{2} \sigma_x + \alpha p \sigma_z+ \frac{\Delta(t)}{2} \sigma_z ,
%\eeqa
%where $\Omega(t)$ is Raman coupling strength.
To drive the state $|n,\uparrow \ra $ (where  $|n,\uparrow \ra = e^{-i  \alpha x} | n \ra |\uparrow \ra$ and $|n \ra$ is given in Eq. \ref{wavefunction}) to the closest one $|l, \downarrow \ra$, where $l=n+1$, see Fig. \ref{schematic},
we construct a $2\times2$ Hamiltonian by taking the matrix elements of $\mathcal{H}_0$ (Eq. \ref{Href}) in the basis of these two states,
and shift it into the symmetric form
%%%%%%%%%%%%%
\beq
\label{H0-nature}
H_0(t) = \frac{1}{2}\left[
\begin{array}{cc}
Z  & X+ i Y
\\
X - i Y & -Z
\end{array}
\right],
\eeq
%%%%%%%%%
where $Z = E_n-E_l+\Delta(t)$ is a time-dependent ``energy gap'', $X = \Omega(t)\textrm{Re}[G]$, $Y = \Omega(t) \textrm{Im}[G]$,  and $G = \la n| e^{2i\alpha x } | l \ra $. The wavefunction corresponding to Hamiltonian of Eq. (\ref{H0-nature}) is of the form $\bm\psi = (\psi_1, \psi_2)^T$, where $\psi_1 = \la n,\uparrow|\bm\psi \ra$ and $\psi_2 = \la l,\downarrow|\bm\psi \ra$.
The dynamical invariant of $H_0$,
\beqa
\label{invariant}
I (t) = \frac{\lambda_0}{2}
\left(
\begin{array}{cc}
\cos{\theta_a} & \sin{\theta_a} e^{i \varphi_a}
\\
\sin{\theta_a} e^{-i \varphi_a} & -\cos{\theta_a}
\end{array}
\right),
\eeqa
where $\lambda_0$ is a constant to keep $I(t)$ with units of energy, is constructed by the, yet unknown, orthogonal eigenstates $|\chi_{\pm} (t) \ra$
%%%%%%%%%%%%%%%%%%%%%%%%%%%%%%
\beqa
|\chi_{+}(t) \ra
&=&
\left(\begin{array}{c}
\cos\displaystyle{\frac{\theta_a}{2}}  e^{i \varphi_a/ 2}
\\
\sin\displaystyle{\frac{\theta_a}{2}}  e^{-i \varphi_a/ 2}
\end{array}
\right),
\\
|\chi_{-}(t)  \ra
&=&
\left(\begin{array}{c}
 \sin \displaystyle{\frac{\theta_a}{2}} e^{-i \varphi_a/ 2}
\\
- \cos \displaystyle{\frac{\theta_a}{2}} e^{i \varphi_a/ 2}
\end{array}
\right),~~~~~
\eeqa
where $\theta_a$ and $\varphi_a$ are the polar and the azimuthal angles for the eigenstates of the invariant. %(with the subscript $a$).
According to the Lewis-Riesenfeld theory, the solution of the Schr\"{o}dinger equation, $i \partial_t \psi=H_0(t) \psi$, is a superposition
of orthonormal ``dynamical modes", $\bm\psi (t) = \sum_n C_n e^{i \xi_n} |\chi_{n} (t) \ra$ \cite{LR}, where $C_n$ are time-independent amplitudes and
the $\xi_n$ Lewis-Riesenfeld phases. Here, we set the trajectory of the actual state evolution along   $|\chi_+ \ra$. From the invariant condition,
\beqa
\frac{dI(t)}{dt} \equiv i \frac{\partial I(t)}{\partial t} - [H_0(t), I(t)] =0,
\eeqa
we find equations for $\theta_a$ and $\varphi_a$,  from which the two controllable parameters are
given by
%%%%%%%%%%%%%%%%%%
\beqa
\label{theta1}
\Omega (t) &=& -\frac{\dot{\theta}_a}{ |G| \sin(\phi - \varphi_a)},
\\
\label{Delta}
\Delta (t) &=& E_l - E_n - \dot{\varphi}_a -\frac{\dot{\theta}_a \cos{\theta_a} \cos (\phi-\varphi_a)}{\sin{\theta_a} \sin(\phi-\varphi_a)},
\eeqa
%%%%%%%%%%%%%%%%%%
where $\phi = \arctan(\textrm{Im}[G]/\textrm{Re}[G])$.
To fulfill the state transfer, $\theta_a$ is set by a polynomial ansatz $\theta_a = \Sigma_{n=0}^3 a_n t^n$ with boundary conditions $\theta_a(0) = 0$, $\theta_a(t_f) = \pi$, $\dot{\theta}_a(0) = \dot{\theta}_a(t_f) = 0$.
These conditions imply the commutativity of the Hamiltonian and the invariant at the boundary times.
Then  $\Delta(t)$ would diverge at $t=0$ and $t=t_f$. To cancel these two singularities, we impose $\dot \theta_a \cot(\phi-\varphi_a) = c \sin \theta_a $, where $c$ is a real number. This results in $\dot{\varphi}_a(0^+) = c$, $\dot{\varphi}_a(t_f^-) = -c$, and leads to $\Delta(0^+) = -E_n +E_l-3c/2$ and $\Delta(t_f^-) = -E_n +E_l+3c/2$.
%Thereby,
%the gap $Z$ of the Hamiltonian $H_0$,
%Eq. (\ref{H0-nature}),  at the initial and final times is small due to small $c$.

\begin{figure}[t]
\scalebox{0.70}[0.70]{\includegraphics{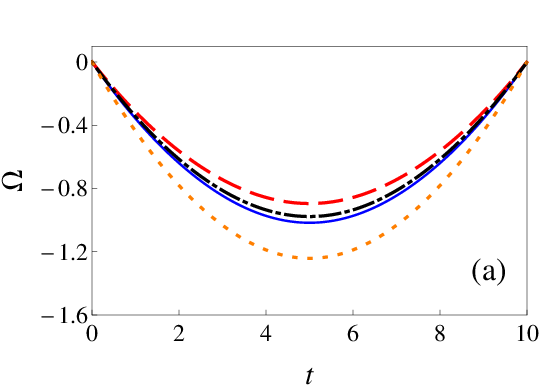}}
\\
\scalebox{0.70}[0.70]{\includegraphics{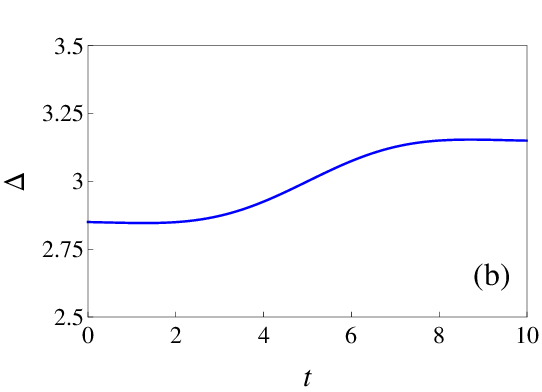}}
\caption{(Color online) (a) Time dependence of Raman coupling strength $\Omega$ with different SO coupling strength $\alpha = 0.8$ (solid, blue), $\alpha = 1.2$ (dashed, red), $\alpha = 1.6$ (dot-dashed, black), $\alpha=2$ (dotted, orange). (b) Time dependence of the detuning $\Delta$, which is independent of $\alpha$. Other parameters are $t_f=10$, $c=0.1$ in both plots. }
\label{F-nature}
\end{figure}

Next, we put forward one example with the Morse potential parameter $A = 8$.
We take $n=0$, $l=1$ and the operation time $t_f=10$. $\Omega$ and $\Delta$ are shown
in Fig. \ref{F-nature} for different values of $\alpha$.
%%%%%%%%%%%%%%%%%%%%%%%%%%%%%%%%%%%%%%%%%%%%%%%%%%%%%%%%%%%%%%%%%%%%%%%%%%%%%%%%%%%%%%%%%%%%%%%%%%%%%%%%%%%%%%%%%
%%%%%%%%%%%%%%%%%%%%%%%%%%%%%%%%%%%%%%%%%%%%%%%%%%%%%%%%%%%%%%%%%%%%
The time-dependent ``energy gap" and detuning $\Delta$ stay unaffected by $\alpha$. This is because $\phi-\varphi$ and $\dot\varphi$ only depend on $\theta_a$. Consequently, to drive the state transition over the same gap with different $\alpha$,  $\Omega(t) G$ should not be changed by $\alpha$. Thus, when increasing $\alpha$ from zero, $|\Omega|$ decreases first and then increases, as shown in Fig. \ref{F-nature} (a).

\begin{figure}[t]
\begin{center}
\scalebox{0.7}[0.7]{\includegraphics{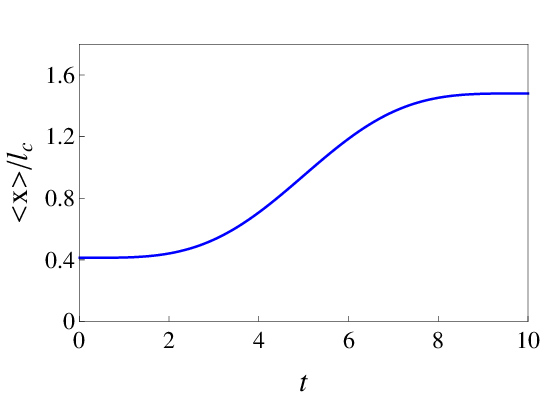}}
\caption{Time dependence of expectation value of coordinate $\la x \ra$ in the unit of $l_\textrm{c}$ for non-interacting atoms, with $t_f = 10$, $\alpha=1.6$, and $c=0.1$.
The initial state is $|0,\uparrow \ra$ and the final one is $|1, \downarrow \ra$.
}
\label{x}
\end{center}
\end{figure}

\begin{figure}[t]
\begin{center}
\scalebox{0.45}[0.45]{\includegraphics{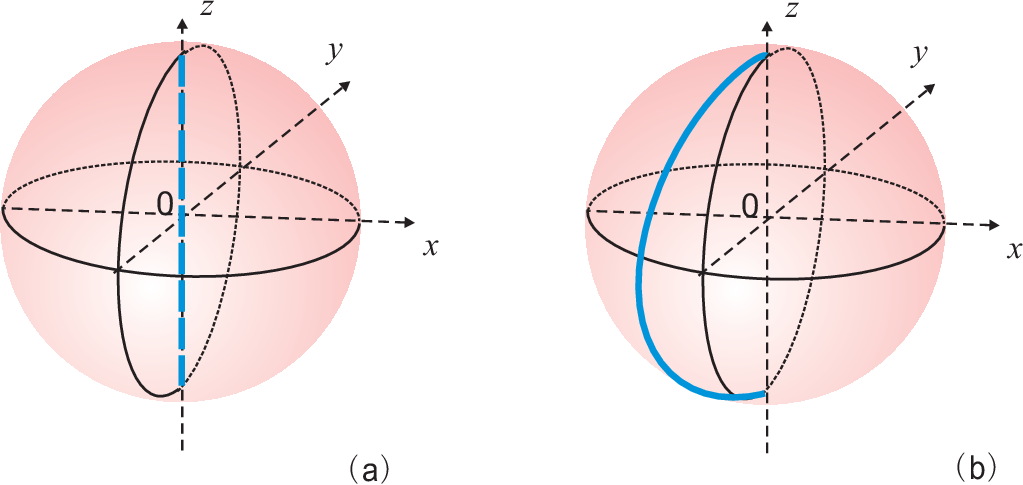}}
\caption{(Color online) Trajectory of state evolution (blue bold line) inside the Bloch sphere in the spin space (a) and on the Bloch sphere in the total space (b).}
\label{trajectory}
\end{center}
\end{figure}

Here we transfer the state from the ground state to the first excited with flipped spin at $t = t_f$. During the transition, the expectation value of the coordinate
\beqa
\la x \ra = \la \bm\psi | \hat{x} | \bm\psi \ra
\eeqa
varies with time. The expectation value of the coordinate at the target state differs from that of the initial state, because of the asymmetry of the Morse potential, see
Fig. \ref{x}.
%%%
%%%
%%%%%%%%%%%%%%%%%%%%%%%%%%%%%%%%%%%%%%%%%%%%%%%%%%%%%%%%%%%%%%%%%%%%%%%%%%%%%%%%%%%%%%%%%%%%%%%%%%%%%%%%%%%%%%%%%%%%%%%%%%%%%%%%%%%%%%%%%%%%%%%%%%%%%%%%%%%%%%%%%%%%%%%%The spin dynamics during the transition can be characterized by a spin polarization
%
\beqa
P_i (t)= \la \bm\psi | \hat{\sigma}_i | \bm\psi \ra,
\eeqa
with components
\beqa
P_x &=& \int_{-\infty}^{+\infty} (\psi_2^* \psi_1 + \psi_1^* \psi_2) dx,
\\
P_y &=& \int_{-\infty}^{+\infty} i(\psi_2^* \psi_1 - \psi_1^* \psi_2) dx,
\\
P_z &=& \int_{-\infty}^{+\infty} (\psi_1^* \psi_1 - \psi_2^* \psi_2) dx.
\eeqa
%
%%%
\begin{figure}[t]
\begin{center}
\scalebox{0.7}[0.7]{\includegraphics{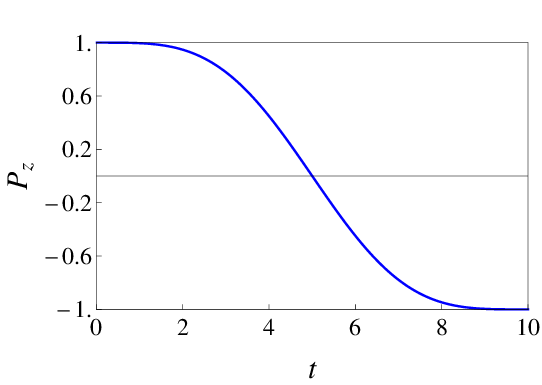}}
\caption{Time evolution of $z$-component of spin polarization $P_z$. The parameters are the same as those in Fig. \ref{x}. }
\label{pz}
\end{center}
\end{figure}
%%%
For initial and final times $P_z(0)=1$ and $P_z(t_f) = -1$, corresponding to spin up and spin down states, respectively, as shown in Fig. \ref{pz}. The spatial orthogonality of eigenstates leads to zero $P_x$, $P_y$ during the operation time, which indicates that the transfer occurs along mixed states ($z$-axis) in spin space, see Fig. \ref{trajectory}. By contrast in the full space the trajectory is a pure state one, along the surface of the
Bloch sphere.

\begin{figure}[t]
\begin{center}
\scalebox{0.65}[0.65]{\includegraphics{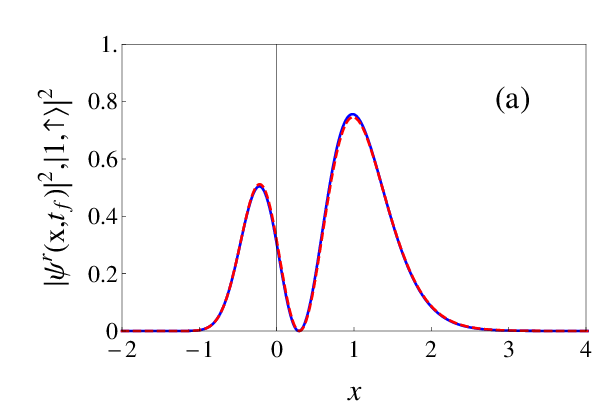}}
\\
\scalebox{0.65}[0.65]{\includegraphics{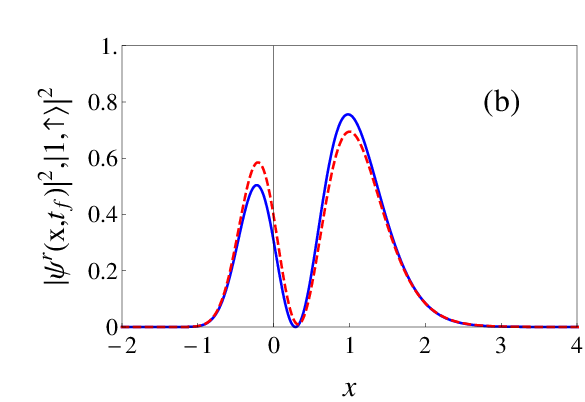}}
\caption{(Color online) Norm square of $|1,\uparrow\ra$ (blue bold line) and $\bm\psi^r(x,t_f)$ (red dashed line) for $c=0.1$ (a) and $c=1.5$ (b).}
\label{norm-square}
\end{center}
\end{figure}

To check the validity of the effective two-level system, we numerically derive the wavefunction $\bm{\psi}^r(x, t)$ by solving the time-dependent Shr\"{o}dinger equation with Hamiltonian Eq. (\ref{Href}) by using $\Delta(t)$ and $\Omega(t)$ inversely designed from Eqs. (\ref{theta1}) and (\ref{Delta}). At the final time, the fidelity $F=|\la l,\downarrow| \bm{\psi}^r(x,t_f) \ra|^2$ is obtained as $0.9966$ for $c=0.1$, corresponding to the small gap $\Delta E=0.15$.
When increasing the energy gap of two-level system, the influence of other states become more pronounced,
so that the fidelity will be decreased. For example, $F=0.979$ for $c = 1.5$, corresponding to $\Delta E =2.25$.
The wavefunction are also compared in Fig. \ref{norm-square} for different energy gap, in which illustrates that
the approximation of a two-level system is valid in the 1D setting for narrow gap between the interest states.

%%%%%%%%%%%%
%%%%%%%%%%%%
%
%
%%%%%%%%%%%%
%%%%%%%%%%%%
%
%
%
%
\section{Scheme 2: State transition by tunable SO coupling direction and time-dependent effective magnetic field}
The state transition from $|n,\uparrow\ra$ to $|l, \downarrow\ra$ can be performed as well
by a Hamiltonian
%
%The Hamiltonian for effective 1D motion of ultracold 2-level atoms confined in a potential in the presence of the external magnetic field and SO coupling is written as
%
\beqa
\label{H-mathcal}
\mathcal{H} = \frac{p^2}{2 } + U(x) + \alpha p (\bm{\sigma} \cdot \bm{n}_1) + \frac{\beta(t)}{2} (\bm{\sigma} \cdot \bm{n}_2),
\eeqa
where $\bm{\sigma}$ is the Pauli matrix vector, and $\beta$ is an effective Zeeman splitting induced by an effective magnetic field in $\bm{n}_2$ direction. The SO coupling and the effective magnetic field are
applied in the directions $\bm{n}_1 = (\sin\theta_1 \cos\varphi_1, \sin\theta_1 \sin\varphi_1, \cos\theta_1)$,
$\bm{n}_2 = (\sin\theta_2 \cos\varphi_2, \sin\theta_2 \sin\varphi_2, \cos\theta_2)$, respectively. The polar and azimuthal angles $\theta_j$ and $\varphi_j$ ($j=1,2$) are tunable parameters.
%%%%%%%%%%%%%%%%%%%%%%%%%%%%%%%%%%%%%%%%%%%%%%%%%%%%%%%%%%
\begin{figure}[t]
\scalebox{0.70}[0.70]{\includegraphics{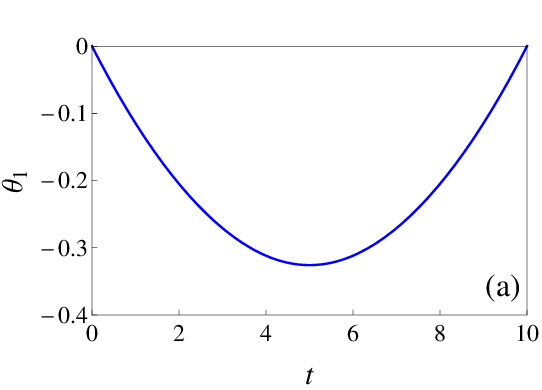}}
\\
\scalebox{0.70}[0.70]{\includegraphics{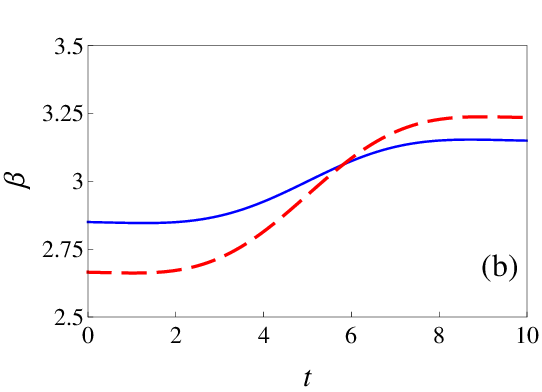}}
\\
\scalebox{0.7}[0.7]{\includegraphics{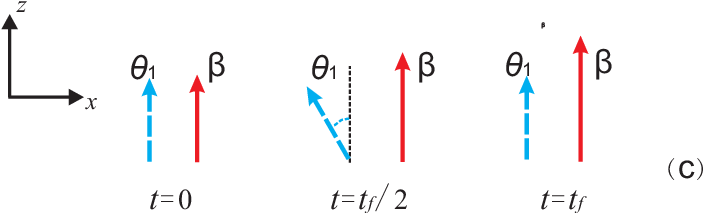}}
\caption{(Color online) (a) Dependence of angle $\theta_1$ of SO coupling field for noninteracting or interacting atoms.
(b) Effective Zeeman splitting $\beta$ versus time $t$,  without interaction (solid, blue), repulsive interaction $g_{11} = 0.3$, $g_{22} = 0.2$, $g_{12}=g_{21}=0.115$ (dashed, red).% and  attractive interaction $g_{11} = -0.3$, $g_{22} = -0.2$, $g_{12}=g_{21}=-0.115$ (dot-dashed, black).
(c) Schematic configuration of the two controllable parameters $\beta$ and $\theta_1$ at $t=0$, $t=t_f/2$ and $t=t_f$. In all the plots, the initial state is $|0,\uparrow \ra$ and the final one is $|1, \downarrow \ra$. Other parameters are $t_f=10$, $\alpha=1.6$, $c=0.1$.
}
\label{control-variables}
\end{figure}
\subsection{non-interacting atoms}
First, we consider non-interacting atoms and construct the $2\times2$ Hamiltonian as a two-level system by taking the matrix elements of $\mathcal{H}$ (Eq. \ref{H-mathcal}) in the basis of $|n,\uparrow \ra$, $|l,\downarrow \ra$,
%
%\begin{widetext}
\beqa
\label{H-TOTAL}
H(t)=
%\displaystyle{\frac{m \alpha^2}{2}} - m \alpha^2 \cos\theta_1 +
\left[
\begin{array}{cc}
 E_n  + \displaystyle{\frac{\beta}{2}} \cos\theta_2  &  M
\\
M^*  & E_l - \displaystyle{\frac{\beta}{2}} \cos\theta_2
\end{array}
\right],
\eeqa
%\end{widetext}
%
where $M=(\alpha^2 G + \alpha K)\sin\theta_1 e^{-i \varphi_1} + \displaystyle{\frac{\beta}{2}} G \sin\theta_2 e^{-i \varphi_2}$,
$K = \la n | e^{2i \alpha x} {p} |l \ra$, and the asterisk represents complex conjugate.
There are many possibilities to perform the state transfer.
%With the application of inverse engineering on the two-level system, only two controllable parameters are necessary to perform the state transition.
As an example, we set $\theta_2 = \varphi_1 = \varphi_2 = 0$, and tune the angle, $\theta_1 \equiv \theta_1(t)$, the direction of SO coupling field, and the effective Zeeman splitting $\beta \equiv \beta(t)$. We approximate $\sin\theta_1 \sim \theta_1$, and $\cos\theta_1 \sim 1- \theta_1^2 / 2$. Symmetrizing the $2 \times 2$ Hamiltonian in Eq. (\ref{H-TOTAL}),
we obtain the reduced form
%%%%%%%%%%%%%
\beq
\label{H0}
\tilde{H}_0(t) = \frac{1}{2}\left[
\begin{array}{cc}
Z  & \tilde{X}+ i \tilde{Y}
\\
\tilde{X} - i \tilde{Y} & -Z
\end{array}
\right],
\eeq
%%%%%%%%%
where $\tilde{X} = 2 \theta_1(t) \textrm{Re}[M]$, $\tilde{Y} = 2 \theta_1(t) \textrm{Im}[M]$.
Two equations for $\theta_a$ and $\varphi_a$  are obtained through the definition of the invariant, Eq. (\ref{invariant}),
%%%%%%%%%%%%%%%%%%
\beqa
\label{theta1}
\theta_1 (t) &=& -\frac{\dot{\theta}_a}{2 |M| \sin(\phi - \varphi_a)},
\\
\label{Delta}
\beta (t) &=& E_l - E_n - \dot{\varphi}_a -\frac{\dot{\theta}_a \cos{\theta_a} \cos (\phi-\varphi_a)}{\sin{\theta_a} \sin(\phi-\varphi_a)},
\eeqa
%%%%%%%%%%%%%%%%%%
where $\phi = \arctan(\textrm{Im}[M]/\textrm{Re}[M])$. The SO coupling strength $\alpha=1.6$ is fixed. With the application of the same ansatzes of $\theta_a$ and $\varphi_a$ used in Sec. III, we find the time-dependence of two controllable variables $\theta_1$ and $\Delta$ in Fig. \ref{control-variables}.
%The maximal absolute value of $\theta_1$ is about $0.32$ at $t = t_f/2$, satisfying the approximation $\theta_1\ll1$. ******REALLY?******
%The magnitude of the effective magnetic field corresponding to $\Delta$ is on the order of mT.
The application of these two parameters is schematically illustrated in Fig. \ref{control-variables} (c).
\subsection{interacting BEC}
Let us now consider a two-component BEC of atoms, (e.g., $^{87}\textrm{Rb}$)
with synthetic SO coupling in an effective magnetic field.
The wavefunction $\bm{\Psi}(x,t) = (\Psi_\uparrow,\Psi_\downarrow)^T$, where $\Psi_\uparrow = \la x, \uparrow| \bm{\Psi} \ra$, $\Psi_\downarrow = \la x, \downarrow| \bm{\Psi} \ra$,
satisfies the coupled Gross-Pitaevskii equations (GPE)
%%%%%%%%%%%%%
%\begin{widetext}
\beqa
\label{GPE}
i\frac{d \Psi_\uparrow}{dt} &=& \Big[\frac{p^2}{2}\! +\! U(x)\! +\! \alpha p \cos\theta_1\! +\! \frac{\beta(t)}{2} \cos\theta_2 \!+\! g_{\uparrow\uparrow} |\Psi_\uparrow |^2
\nonumber\\
&+&\!g_{\uparrow\downarrow} |\Psi_\downarrow |^2 \Big] \Psi_\uparrow\! +\!\! \left(\!\alpha p \sin\theta_1 e^{-i \varphi_1} \!+\! \frac{\beta}{2} \sin\theta_2 e^{-i \varphi_2}\! \right)\! \Psi_\downarrow,
\nonumber
\\
i\frac{d \Psi_\downarrow}{dt} &=& \Big[\frac{p^2}{2} + U(x) \!-\! \alpha p \cos\theta_1 \!-\! \frac{\beta(t)}{2} \cos\theta_2 \!+\! g_{\downarrow\uparrow} |\Psi_\uparrow |^2
\nonumber\\
&+& g_{\downarrow\downarrow} |\Psi_\downarrow |^2 \Big] \Psi_\downarrow \!+\! \left(\alpha p \sin\theta_1 e^{i \varphi_1} \!+\! \frac{\beta}{2} \sin\theta_2 e^{i \varphi_2}\right) \Psi_\uparrow.
\nonumber\\
\eeqa
%\end{widetext}
%%%%%%%%%
The intra-component and inter-component atomic interaction constants are represented as $g_{jj}$, and $g_{jk}$ ($j \neq k =\uparrow,\downarrow$), respectively.
As for the non-interacting gas we also take $\theta_1$ and $\beta$ as controllable variables. (The scheme
in Sec. II can be generalized similarly.) The inverse engineering based on a Lewis-Riesenfeld invariant, however, is not applicable, as Eq. (\ref{invariant}) is not the invariant of the new Hamiltonian $H_1$,
%%%%%%%%%%%%%%%%%%
\beqa
\label{H1}
H_1 &=&\tilde{H}_0
\nonumber
\\
&+& \left[
\begin{array}{cc}
g_{11} |\psi_1|^2 + g_{12} |\psi_2|^2 & 0
\\
0 & g_{21} |\psi_1|^2 + g_{22} |\psi_2|^2
\end{array}\right]\!\!.
\eeqa
%%%%%%%%%%%%%%%%%%
The wavefunction can be normalized by the factor $N = \int_{-\infty}^\infty |\bm \Psi|^2 dx$. The effective factors of the intra- and the inter-components are
%%%%%%%%%%%%%%%%%%
\beqa
\label{effective-g}
\nonumber
g_{11} &=& g_{\uparrow\uparrow} \textrm{Q}(0,0),  \quad g_{22} = g_{\downarrow\downarrow} \textrm{Q}(1,1),
\\
\nonumber
g_{12} &=& g_{\uparrow\downarrow} \textrm{Q}(0,1), \quad g_{21} = g_{\downarrow\uparrow} \textrm{Q}(0,1).
\eeqa
where
\beq
\label{integral}
%\textrm{I}(n) &=& \int_{-\infty}^\infty |\la x | n \ra|^4 dx,
%\\
\textrm{Q}(n,l) = \int_{-\infty}^\infty |\la x | n \ra|^2 |\la x | l \ra|^2 dx,
\eeq
and $n$, $l$ are the quantum numbers of the orbital states.
From the wavefucntion Eq. (\ref{wavefunction}), we can calculate the ratios $ \textrm{Q}(0,0) / \textrm{Q}(1,1) = 1.5$, $ \textrm{Q}(0,1)/[\textrm{Q}(0,0)+\textrm{Q}(1,1)] = 0.23$. By setting the same value for $g_{\uparrow\uparrow}$, $g_{\uparrow\downarrow}$, $g_{\downarrow\uparrow}$ and $g_{\downarrow\downarrow}$, we finally determine $g_{11}$, $g_{22}$, $g_{12}$ and $g_{21}$.

Inverse engineering is still feasible for the nonlinear system by means of the state ansatz
\beqa
\label{PSI}
\bm{\psi}(t)= \left(\begin{array}{c}
\cos\displaystyle{\frac{\theta_p}{2}} e^{i \varphi_p/2}
\\ \sin\displaystyle{\frac{\theta_p}{2}} e^{-i \varphi_p/2}
\end{array}\right) e^{i\gamma},
\eeqa
where $e^{i\gamma}$ is the global phase.
%Note that $\theta_p$ and $\varphi_p$ are the polar and azimuthal angles of the trajectory of the state evolving on the Bloch sphere, different from the previous auxiliary angles $\theta_a$ and $\varphi_a$.

Choosing for $\theta_p$ and $\varphi_p$ the same ansatz used before for
$\theta_a$ and $\varphi_a$, we get from the GPE
%ion in Eq. (\ref{PSI}), we can get equations for $\theta_p$ and $\varphi_p$, with the structure of those for . , we find the expressions for $\theta_1$ and $\beta$,
%%%%%%%%%%%%
\beqa
\label{theta1BEC}
\theta_1 (t) &=& -\frac{\dot{\theta}_p}{2 |M| \sin(\phi - \varphi_p)},
\\
\label{DeltaBEC}
\beta(t) &= & E_l - E_n -\dot{\varphi}_p -\frac{\dot{\theta}_p \cos{\theta_p} \cos (\phi-\varphi_p)}{\sin{\theta_p} \sin(\phi-\varphi_p)}
\nonumber
\\
 &&- g_{11} \cos^2\frac{\theta_p}{2} - g_{12} \sin^2\frac{\theta_p}{2}
\nonumber\\
 &&+ g_{22}\sin^2\frac{\theta_p}{2} + g_{21}\cos^2\frac{\theta_p}{2}.
\eeqa
%%%%%%%%%%%%%%%%%%
The function $\theta_1$ in Eq. (\ref{theta1BEC}) keeps
the same form as that without the interaction (Eq. (\ref{theta1})), shown in Fig. \ref{control-variables} (a). In contrast,
the interaction in diagonal terms results in changes in the amplitude of the external effective magnetic field, as seen in Fig. \ref{control-variables} (b).
In other words, the modulation in $\beta$ compensates the non-linear terms.
\section{Robustness}
%
%
%
%Our protocols provide a way to inversely design the controllable parameters for state transfer. Actually, the time evolution  of matrix elements in $H_0$ and $\tilde{H}_0$ are the same, although they are implemented by different physical parameters.
%******NOT CLEAR IF WE SHOULD PUT ATTRACTIVE CASE******
We shall now test the stability of the protocol
based on the Hamiltonian (\ref{H1}) with respect to fluctuations of the effective magnetic
field caused by systematic errors and noise, which induce shifts of the diagonal terms.

We first consider that the effective Zeeman splitting, deviates from the theoretical value as  $\beta^{real} =  \beta(1+\lambda)$, where $\lambda$ is constant, but generally unknown.
%%%
\begin{figure}
\begin{center}
\scalebox{0.7}[0.7]{\includegraphics{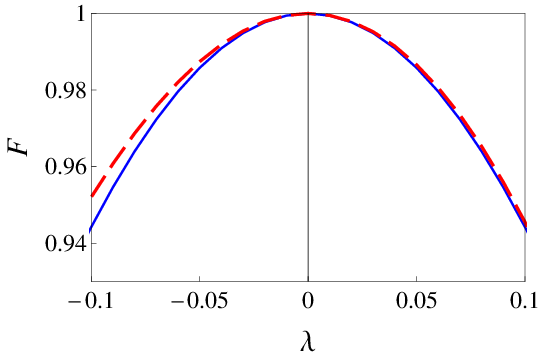}}
\caption{(Color Online) Fidelity with respect to the relative error of the magnetic field $\lambda$ for condensates with non-interaction (solid, blue),
the repulsive interaction $g_{11}=0.3$, $g_{22} =0.2$, $g_{12}=g_{21}=0.115$ (dashed, red).
%, and attractive interaction $g_{11}=-0.3$, $g_{22} =-0.2$, $g_{12}=g_{21}=-0.115$ (dot-dashed, black).
}
\label{systematic}
\end{center}
\end{figure}
%%%
The fidelity $F = |\la l,\downarrow | \bm\psi(t_f) \ra|^2$ with respect to $\lambda$ is compared for  non-interaction, and for attractive and repulsive interactions in Fig. \ref{systematic}, showing the stability
around $\lambda=0$.
%As the repulsive interaction decreases $\beta$  at initial time, in this case the system gains less influence from the same deviation percentage.
%As a result, the fidelity with repulsive interaction is higher. This also explains for the attractive interaction, fidelity decreases, although in all three cases, fidelity is already high. In other words, repulsive interaction can decreases the influence from the systematic errors.

\begin{figure}[t]
\begin{center}
\scalebox{0.7}[0.7]{\includegraphics{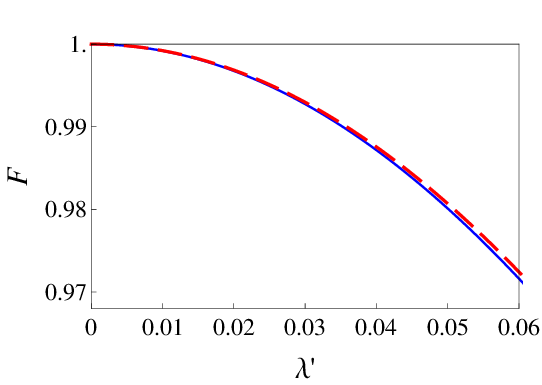}}
\caption{(Color Online) Fidelity with respect to $\lambda'$ with the comparisons of non-interaction (solid, blue), the
repulsive interaction $g_{11}=0.3$, $g_{22} =0.2$, $g_{12} = g_{21} =0.115$ (dashed, red).
%and attractive interaction $g_{11}=-0.3$, $g_{22} =-0.2$, $g_{12}=g_{21}=-0.115$ (dot-dashed, black).  $t_f=10$, $\alpha=1.6$, $c=0.1$.
}
\label{amplitude}
\end{center}
\end{figure}

Besides the systematic errors, we also consider a noisy perturbation, i.e. the Hamiltonian $H_1$ (Eq. (\ref{H1})) perturbed by a stochastic part $H^n$. The GPE is modified as
%%%%%%%%%%%%
\beqa
\label{GPE-stochastic}
i\frac{d \bm{\psi}(t)}{dt} = (H_1+H^n)\bm{\psi}(t),
\eeqa
%%%%%%%%%%%%%%%%%%
where $H^n = \lambda' H' \xi(t)$, $ \la \xi(t) \ra = 0$, $\la \xi(t) \xi(t') \ra = \delta(t-t')$, $\lambda'$ is the noise strength, and
$H'$ is
\beqa
\label{Hn}
H' = \frac{1}{2}\left[
\begin{array}{cc}
\beta& 0
\\
0 & \beta
\end{array}\right].
\eeqa
The density matrix obeys now \cite{noise}
\beqa
\label{master-noise-bloch}
\dot\rho = - i [H_1, \rho] - \frac{\lambda'^2}{2}[H',[H',\rho]].
\eeqa
We introduce the Bloch vector with components $u = \rho_{1-1} + \rho_{-11}$, $v = -i(\rho_{1-1} - \rho_{-11})$, and $w = \rho_{11} - \rho_{-1-1}$, and obtain
\beqa
\label{master-noise}
\nonumber
\dot{u} &=& -\frac{1}{2} \lambda'^2 \Delta^2 u + (Z+g_{d} + g_s w - g_s' w + g_d') v - Y w,
\\
\nonumber
\dot{v} &=& (-Z -g_{d} - g_s w - g_d' + g_s' w) u -\frac{1}{2} \lambda'^2 \Delta^2 v + X w,
\\
\dot{w} &=& Y u - X v,
\eeqa
where $g_d = ({g}_{11} - g_{22})/2$, $g_s = ({g}_{11} + g_{22})/2$, $g_d' = (g_{12} - g_{21})/2$, $g_s' = (g_{12} + g_{21})/2$.
We calculate numerically the fidelity in Fig. \ref{amplitude} which, again, shows stability around
$\lambda=0$.

For both types of perturbations the stability region where the fidelity is close enough to one
may be broadened by inverse
engineering the process as in \cite{noise}.
%
%Compared with systematic errors, for the larger amplitude of the magnetic field, amplitude noise has stronger influence. For example, as shown in Fig. \ref{amplitude},
%the repulsive interaction corresponding to the largest field amplitude has the lowest fidelity, although the difference from the non-interaction, the attractive interaction is almost not distinguishable.
%%%
%%%
%
%
\section{Summary}
Schemes for the state transfer of SO-coupled bosonic atoms trapped in a Morse potential are proposed.
By invariant-based inverse engineering, for non-interacting atoms, time-dependence of Raman coupling and detuning are designed to transfer internal and orbital states. The transition is coherent but occurs via
mixed-states in pseudo-spin space.
It also induces a displacement of the atomic wave due to the asymmetry of Morse potential.
%Meanwhile, mixed states in the pseudo-spin space show up.
An alternative scheme is to tune the SO coupling direction and the amplitude of the effective magnetic field.
For the interacting BEC, the amplitude of the effective magnetic field can be designed to compensate the non-linearity. The protocols are stable with respect to
systematic errors and amplitude noise in the applied effective magnetic field.
Similar ideas may be applied to design fast transitions between a bound state and the continuum.
\section*{Acknowledgement}
This work is partially supported by the NSFC (Grant Nos. 11474193, 61404079, and 61176118), the Shanghai Municipal Science and Technology Commission (Grant Nos.
13PJ1403000 and 14YF1408400), the Specialized Research Fund for the Doctoral Program (Grant No. 2013310811003), the Program for Eastern Scholar, the Basque Country
Government (Grants Nos. IT472-10), Ministerio de Econom\'{i}a y Competitividad (Grant
No. FIS2012-36673-C03-01), and UPV/EHU program UFI
11/55.


\section*{References}
\begin{thebibliography}{10}
%\bibitem{Bowler} R. Bowler, J. Gaebler, Y. Lin, T. R. Tan, D. Hanneke, J. D. Jost, J. P. Home, D. Leibfried, and D. J. Wineland, Phys. Rev. Lett. \textbf{109}, 080502 (2012).

%\bibitem{Walther} A. Walther, F. Ziesel, T. Ruster, S. T. Dawkins, K. Ott, M. Hettrich, K. Singer, F. Schmidt-Kaler, and U. Poschinger, Phys. Rev. Lett. \textbf{109}, 080501 (2012).


\bibitem{Schmiedmayer-JPB} R. B\"{u}cker, T. Berrada, S. van Frank, J.-F. Schaff, T. Schumm, J. Schmiedmayer, G. J\"{a}ger, J. Grond, and U. Hohenester, J. Phys. B: At. Mol. Opt. Phys. \textbf{46}, 104012 (2013).


\bibitem{multiplexing} S. Mart\'{i}nez-Garaot, E. Torrontegui, X. Chen, M. Modugno, D. Gu\'{e}ry-Odelin, S. Y. Tseng, and J. G. Muga, Phys. Rev. Lett. \textbf{111}, 213001 (2013).

\bibitem{SOC-tunneling} Y. Ban and E. Y. Sherman, Phys. Rev. A \textbf{85}, 052130 (2012).

\bibitem{SOC-measurement} D. Sokolovski and E. Ya. Sherman, Phys. Rev. A \textbf{89}, 043614 (2014).



\bibitem{SOC-BEC1} T. D. Stanescu, B. Anderson, and V. Galitski, Phys. Rev. A \textbf{78}, 023616 (2008).

\bibitem{SOC-BEC2} Y. J. Lin, K. Jim\'{i}nez-Garc\'{i}a and I. B. Spielman, Nat. \textbf{471}, 83 (2011).

\bibitem{SOC-fermion1} X.-J. Liu, M. F. Borunda, X. Liu, and J. Sinova, Phys. Rev. Lett. \textbf{102}, 046402 (2009).

\bibitem{SOC-fermion2} P. Wang, Z.-Q. Yu, Z. Fu, J. Miao, L. Huang, S. Chai, H. Zhai, and J. Zhang, Phys. Rev. Lett. \textbf{109}, 095301 (2012).

\bibitem{SOC-fermion3} L. W. Cheuk, A. T. Sommer, Z. Hadzibabic, T. Yefsah, W. S. Bakr, and M. W. Zwierlein, Phys. Rev. Lett. \textbf{109}, 095302 (2012).

\bibitem{review1} J. Dalibard, F. Gerbier, G. Juzeliunas, and P. Ohberg, Rev. Mod. Phys. 83, 1523 (2011).
\bibitem{review2} H. Zhai, Int. J. Mod. Phys. \textbf{26}, 1230001 (2012).
\bibitem{review3} V. Galitski and I. B. Spielman, Nature \textbf{494}, 49 (2013).

\bibitem{spin-relaxation-K} T. Yu and M. W. Wu, Phys. Rev. A \textbf{88}, 043634 (2013).

\bibitem{Zitterbewegung-spinhalf} C. L. Qu, C. Hamner, M. Gong, C.W. Zhang, and P. Engels, Phys. Rev. A \textbf{88}, 021604 (2013).

\bibitem{Zitterbewegung-spin1} Y. C. Zhang, S. W. Song, C. F. Liu, and W. M. Liu, Phys. Rev. A \textbf{87}, 023612 (2013).

%\bibitem{SOC-topological} X.-J. Liu, Z.-X. Liu, and M. Cheng, Phys. Rev. Lett. \textbf{110}, 076401 (2013).

\bibitem{Landau-Zener-SOC} A. J. Olson, S. J. Wang, R. J. Niffenegger, C. H. Li, C. H. Greene, and Y. P. Chen, Phys. Rev. A \textbf{90}, 013616 (2014).


\bibitem{PRL104} X. Chen, A. Ruschhaupt, S. Schmidt, A. del Campo, D. Gu\'{e}ry-Odelin, and J. G. Muga, Phys. Rev. Lett. \textbf{104}, 063002 (2010).


\bibitem{STA} E. Torrontegui, S. Ib\'{a}\~{n}ez, S. Mart\'{i}nez-Garaot, M. Modugno, A. del Campo, D. Gu\'{e}y-Odelin, A. Ruschhaupt, X. Chen, and J. G. Muga, Adv. At. Mol. Opt. Phys. \textbf{62}, 117 (2013).

%\bibitem{BEC-transport} E. Torrontegui, X. Chen, M. Modugno, S. Schmidt, A. Ruschhaupt, and J. G. Muga, New J. Phys. \textbf{14}, 013031 (2012).
\bibitem{BanPRL} Y. Ban, X. Chen, E. Ya. Sherman, and J. G. Muga, Phys. Rev. Lett. \textbf{109}, 206602 (2012).

\bibitem{Morse} P. M. Morse, Phys. Rev. \textbf{34}, 57 (1929).

\bibitem{morse-potential} Y. Colombe, et al., J. Opt. B \textbf{5} S155 (2003).

\bibitem{LR} H. R. Lewis and W. B. Riesenfeld, \textit{J. Math. Phys.} \textbf{10}, 1458 (1969).

\bibitem{noise} A. Ruschhaupt, X. Chen, D. Alonso and J. G. Muga, New J. Phys. \textbf{14}, 093040 (2012).

%\bibitem{Schmiedmayer-arxiv}  S. van Frank, A. Negretti, T. Berrada, R. B\"{u}cker, S. Montangero, J.-F. Schaff, T. Schumm, T. Calarco, and J. Schmiedmayer, arXiv:1402.0377.

%\bibitem{Sarandy} M. S. Sarandy, E. I. Duzzioni, and R. M. Serra, 2011, Phys. Lett. A \textbf{375}, 3343 (2011).
%\bibitem{Campo1} A. del Campo, M. M. Rams, and W. H. Zurek, Phys. Rev. Lett. \textbf{109}, 115703 (2012).
%\bibitem{Campo2} A. del Campo, Phys. Rev. Lett. \textbf{111}, 100502 (2013).
%\bibitem{BEC-ff} S. Masuda, and S. A. Rice, Phys. Rev. A \textbf{89}, 033621 (2014).

%\bibitem{BEC-application} J. Est\`{e}ve, J.-B. Trebbia, T. Schumm, A. Aspect, C. I. Westbrook, and I. Bouchoule, Phys. Rev. Lett. \textbf{96}, 130403 (2006); B. V. Hall, S. Whitlock, R. Anderson, P. Hannaford, and A. I. Sidorov, Phys. Rev. Lett. \textbf{98}, 030402 (2007); U. Hohenester, P. K. Rekdal, A. Borzi, and J. Schmiedmayer, Phys. Rev. A \textbf{75}, 023602 (2007); J. Esteve, C. Gross, A. Weller, S. Giovanazzi, and M. K. Oberthaler, Nature (London) \textbf{455}, 1216 (2008).

%\bibitem{double1} Q. Niu, X. G. Zhao, G. A. Georgakis, and M. G. Raizen, Phys. Rev. Lett. \textbf{76}, 4504 (1996).
%\bibitem{Qiu} J. Liu, B. Wu, and Q. Niu, Phys. Rev. Lett. \textbf{90}, 170404 (2003).
%\bibitem{Chen} Y.-A. Chen, S. D. Huber, S. Trotzky, I. Bloch, E. Altman, Nat. Phys. \textbf{7}, 61 (2011).

%\bibitem{triple1} E. M. Graefe, H. J. Korsch, and D. Witthaut, Phys. Rev. A \textbf{73}, 013617 (2006).
%\bibitem{Liu} G. F. Wang, D. F. Ye, L. B. Fu, X. Z. Chen, and J. Liu, Phys. Rev. A \textbf{74}, 033414 (2006).



%shortcut application
%\bibitem{BEC}J. G. Muga, X. Chen, A. Ruschhaup, and D. Gu\'{e}ry-Odelin, J. Phys. B: At. Mol. Opt. Phys. \textbf{42}, 241001 (2009).
%\bibitem{Erik} E. Torrontegui, X. Chen, M. Modugno, A. Ruschhaupt, D. Gu\'{e}ry-Odelin and J. G. Muga, Phys. Rev. A \textbf{85}, 033605 (2012).

%\bibitem{transport1} E. Torrontegui, S. Ib\'{a}\~{n}ez, X. Chen, A. Ruschhaupt, D. Gu\'{e}ry-Odelin, and J. G. Muga, Phys. Rev. A \textbf{83}, 013415 (2011).
%\bibitem{transport2}E. Torrontegui, X. Chen, M. Modugno, S. Schmidt, A. Ruschhaupt, and J. G. Muga, New J. Phys. \textbf{14}, 0130312 (2012).

%\bibitem{Nice} J. F. Schaff, X.-L. Song, P. Vignolo, and G. Labeyrie, Phys. Rev. A \textbf{82}, 033430 (2010).
%\bibitem{Nice2}J. F. Schaff, X. L. Song, P. Capuzzi, P. Vignolo, and G. Labeyrie, EPL \textbf{93}, 23001 (2011).
%\bibitem{Oliver} M. G. Bason, M. Viteau, N. Malossi, P. Huillery, E. Arimondo, D. Ciampini, R. Fazio, V. Giovannetti, R. Mannella, and O. Morsch, Nat. Phys. \textbf{8}, 147 (2012).
%\bibitem{Schmiedmayer} W. Rohringer, D. Fischer, F. Steiner, I. E Mazets, J. Schmiedmayer, and M. Trupke, arXiv:1312.5948.



\end{thebibliography}
\end{document}